\documentclass[preprint,aps,showpacs,preprint,floatfix]{revtex4-1}
\usepackage{graphicx}
\usepackage{subfigure}
\usepackage{array}
\usepackage{color}
\usepackage{amsmath}
\usepackage{amsxtra}
\usepackage{tablists}
\usepackage{diagbox}
\usepackage{amstext}
\usepackage{amssymb}
\usepackage{latexsym}
\usepackage{dsfont}
\usepackage{verbatim}
\usepackage{mathrsfs}
\usepackage{graphicx}
\usepackage[position=t,singlelinecheck=off]{subfig}
\usepackage{caption}
\usepackage{multirow}
\usepackage{marvosym}
\usepackage{floatrow}
\usepackage[utf8]{inputenc}
\usepackage{caption}
\newfloatcommand{capbtabbox}{table}[][\FBwidth]

\newcommand{\beq}{\begin{equation}}
\newcommand{\eeq}{\end{equation}}
\newcommand{\bea}{\begin{eqnarray}}
\newcommand{\eea}{\end{eqnarray}}

\newcommand{\Dfb}{\mbox{$\raisebox{2mm}{\boldmath ${}^\leftrightarrow$}\hspace{-4mm} D$}}

\begin{document}

\title{
Anomalous $H\to ZZ \to 4\ell$ decay and its interference effects at the LHC
}
\author{Hua-Rong He$^1$\footnote{hehr@snnu.edu.cn}, Xia Wan$^1$\footnote{wanxia@snnu.edu.cn}, You-Kai Wang$^1$\footnote{wangyk@snnu.edu.cn}}
\affiliation{$^1$School of physics and Information Technology, Shaanxi Normal University,
Xi'an 710119, China}

\begin{abstract}
We calculate the spinor helicity amplitudes of anomalous $H\to ZZ \to 4\ell$ decay.
After embedding these analytic formulas into the $\texttt{MCFM}$ package,
we study the interference effects between the anomalous $gg\to H\to ZZ \to 4\ell$
process and the SM processes, which are indispensable in the Higgs off-shell region.
Subsequently, the constraints on the anomalous couplings
are estimated using LHC experimental data.

\end{abstract}

\maketitle

\section{Introduction\label{section:introduction}}

Since the 125~GeV/$c^2$ Higgs boson was discovered at the Large Hadron Collider (LHC) in 2012~\cite{Chatrchyan:2012xdj,Aad:2012tfa}, its properties have been tested more and more precisely~\cite{Sirunyan:2018koj,Tao:2018zeu,Magnan:2018spp}.
Even though no new physics
beyond Standard Model (SM) has been confirmed so far,
it is still necessary and meaningful to
search for new physics. 
In this paper we study the anomalous $HZZ$ couplings.

The new physics beyond the SM in the SM effective field theory (SMEFT)
is shown as higher-dimensional operators in the Lagrangian which later supply non-SM interactions.
In this analysis we note these non-SM $HVV$ ($V$ represents $Z,W,\gamma$)
 interactions from 
six-dimensional operators as anomalous $HVV$ couplings, and consider 
them separately from SM loop contributions.  
To scrutinize the Lorentz structures from several anomalous couplings, we calculate 
the scattering amplitudes in the spinor helicity method,
and the analytic formulas are shown symmetrically and elegantly 
in the spinor notations. 

$HVV$ couplings can be probed at the LHC through 
processes including $V^\ast\to VH$ or $H\to VV$ decays.
Among these processes, 
the $gg\to H\to ZZ \to 4\ell$ process,
which is called the golden channel, is the most precise 
and has been studied extensively
in both theoretical studies
~\cite{Gao:2010qx,Bolognesi:2012mm,Anderson:2013afp,
Chen:2012jy,Chen:2013ejz,Chen:2014pia,Chen:2014hqs,Nelson:1986ki,Soni:1993jc,
Chang:1993jy,Arens:1994wd,Choi:2002jk,Buszello:2002uu,
Godbole:2007cn,Kovalchuk:2008zz,Cao:2009ah,DeRujula:2010ys,
Gainer:2011xz,Coleppa:2012eh,Stolarski:2012ps,Boughezal:2012tz,
Avery:2012um,Campbell:2012ct,Campbell:2012cz,Modak:2013sb,Sun:2013yra,
Gainer:2013rxa,Buchalla:2013mpa,
Chen:2013waa,Kauer:2012hd,
Beneke:2014sba,Falkowski:2014ffa,Modak:2014zca,Gonzalez-Alonso:2014rla,
Belyaev:2015xwa,
Gainer:2018qjm}
and experiments at LHC
~\cite{Chatrchyan:2012jja,Khachatryan:2014kca,
Khachatryan:2014iha,
deFlorian:2016spz,Sirunyan:2017tqd,Sirunyan:2019twz,
Aad:2015xua,Aaboud:2018wps,
Aaboud:2018puo}.
Thus, we also choose this golden channel to study anomalous $HVV$ couplings.
To reach a more precise result, both on-shell 
and off-shell Higgs regions can be exploited. 
At the same time, the interference effects between this process and the SM processes
should be included. Especially in the off-shell Higgs region, the 
interference between this process and the continuum process $gg\to ZZ \to 4\ell$
should not be ignored~\cite{Campbell:2013una,Campbell:2011cu}.      
Based on a modified $\texttt{MCFM}$~\cite{Campbell:2013una, Ellis:2014yca} package with anomalous $HZZ$ couplings, we study the interference effects quantitatively. 
Furthermore, we estimate the constraints on the anomalous coupling using CMS 
experimental data at LHC.

The rest of the paper is organized as follows. In Section~\ref{section:theory_calc}, the spinor helicity amplitudes with anomalous couplings are calculated. In Section~\ref{section:numerical result}, the analytic formulas are embedded into the \texttt{MCFM8.0} package and the cross sections for proton - proton collision, especially the interference effects, are shown numerically.
In Section ~\ref{section:constraints}, the constraints on the $HZZ$ anomalous couplings are estimated.
Section~\ref{section:conclusion} is the conclusion and discussion.

\section{theoretical calculation\label{section:theory_calc}}

In this section firstly we introduce the $HZZ$ anomalous couplings, and then we calculate the spinor helicity amplitudes.

\subsection{$HZZ$ anomalous couplings}

In the SM effective field theory~\cite{Buchmuller:1985jz, Grzadkowski:2010es}
the complete form of higher-dimensional operators can be written as
\beq
\mathcal{L}=\mathcal{L}_{SM}+\frac{1}{\Lambda}\sum_k C_k^{5}\mathcal{O}_k^{5}
+\frac{1}{\Lambda^2}\sum_k C_k^{6}\mathcal{O}_k^{6}+\mathcal{O}(\frac{1}{\Lambda^3})~,
\label{complete_form}
\eeq
where $\Lambda$ is the new physics energy scale, and $C_k^{i}$ with $i=5,6$
are Wilson loop coefficients.
As the dimension-five operators $\mathcal{O}_k^{5}$ have no contribution
 to anomalous $HZZ$ couplings, the dimension-six operators $\mathcal{O}_k^{6}$ have leading contributions.
The relative dimension-six operators in the Warsaw basis~\cite{Grzadkowski:2010es} are
\bea
&&\mathcal{O}^6_{\Phi D} =(\Phi^{\dagger}D^{\mu}\Phi)^{\ast}(\Phi^{\dagger}D^{\mu}\Phi), \nonumber \\
&&\mathcal{O}^6_{\Phi W}=
\Phi^\dagger \Phi W^{I}_{\mu\nu}W^{I\mu\nu},~~
\mathcal{O}^6_{\Phi B}= \Phi^\dagger\Phi B_{\mu\nu}B^{\mu\nu},~~
\mathcal{O}^6_{\Phi WB}= \Phi^\dagger \tau^I \Phi W^{I}_{\mu\nu}B^{\mu\nu},
\nonumber \\
&& \mathcal{O}^6_{\Phi \tilde{W}}= \Phi^\dagger\Phi \tilde{W}^{I}_{\mu\nu}W^{I\mu\nu},
~~\mathcal{O}^6_{\Phi \tilde{B}}= \Phi^\dagger\Phi \tilde{B}_{\mu\nu}B^{\mu\nu},  
~~\mathcal{O}^6_{\Phi \tilde{W}B}= \Phi^\dagger \tau^I \Phi \tilde{W}^{I}_{\mu\nu}B^{\mu\nu},
\label{operatorphiw}
\eea
where $\Phi$ is a doublet representation under the $SU(2)_L$ group and the aforementioned Higgs field $H$ is one of its four components;
$D_\mu=\partial_\mu-i g W^{I}_{\mu}T^{I}-ig^{\prime}YB_\mu$, where $g$ and $g^\prime$ are coupling constants, $T^{I}=\tau^{I}/2$, where $\tau^{I}$ are Pauli matrices, $Y$ is the $U(1)_Y$ generator;
$W^{I}_{\mu\nu}=\partial_\mu W^{I}_{\nu}-\partial_\nu W^{I}_\mu-g\epsilon^{IJK}W^{J}_\mu W^{K}_\nu$,
$B_{\mu\nu}=\partial_\mu B_\nu-\partial_\nu B_\mu$,
$\tilde{W}^{I}_{\mu\nu}= \frac{1}{2}\epsilon_{\mu\nu\rho\sigma}W^
{I\rho\sigma}$,
$\tilde{B}_{\mu\nu}=\frac{1}{2}\epsilon_{\mu\nu\rho\sigma}B^
{\rho\sigma}$. 

For the $H\to 4\ell$ process that we are going to take to constrain the 
anomalous $HZZ$ couplings numerically, there are dimension-six operators include $HZ\ell\ell$ contact interaction ~\cite{Barklow:2017awn,Cohen:2016bsd} that can also contribute non-SM effects, 
which are 
\beq
\mathcal{O}^6_{\Phi L}= (\Phi^\dagger \Dfb_\mu\Phi)( \bar{L}\gamma_\mu L),
~~\mathcal{O}^6_{\Phi LT}= (\Phi^\dagger T^I \Dfb_\mu \Phi)( \bar{L}\gamma_\mu T^I L),
~~\mathcal{O}^6_{\Phi e}= (\Phi^\dagger \Dfb_\mu\Phi)(\bar{e}\gamma_\mu e),
\label{operatorphil}
\eeq
where
$\Phi^\dagger \Dfb_\mu\Phi=\Phi^\dagger D_\mu\Phi-D_\mu\Phi^\dagger\Phi$,
~~$\Phi^\dagger T^I\Dfb_\mu\Phi=\Phi^\dagger T^I D_\mu\Phi-D_\mu\Phi^\dagger T^I \Phi$,
$L$,$e$ represent left- and right-handed charged leptons.
One may worry about the pollution caused by the $HZ\ell\ell$ contact interaction from these operators to the $4\ell$ final state when probing  $HZZ$ couplings. 
Nevertheless, we can use certain additional methods to distinguish them. In the off-shell Higgs region, the on-shell $Z$ boson selection
cut can reduce much of the $HZ\ell\ell$ background. In on-shell Higgs region,
the non-leptonic $Z$ decay channel can also be adopted in constraining $HZZ$ couplings. These discussions are not the focus of the current paper and we are not going to examine them in detail here.

After spontaneous symmetry breaking, we get the anomalous $HZZ$ interactions
\beq
\mathcal{L}_{a}=\frac{a_1}{v}M_Z^2 HZ^{\mu}Z_{\mu}-\frac{a_2}{v}HZ^{\mu\nu}Z_{\mu\nu}
-\frac{a_3}{v}HZ^{\mu\nu}\tilde{Z}_{\mu\nu}~,
\label{lagrangian}
\eeq
with 

\bea
a_1&=& \frac{v^2}{\Lambda^2}C^6_{\Phi D}, \nonumber \\
a_2&=&-\frac{v^2}{\Lambda^2}(C^6_{\Phi W}c^2+C^6_{\Phi B}s^2+C^6_{\Phi WB}cs), \nonumber \\
a_3&=&-\frac{v^2}{\Lambda^2}(C^6_{\Phi \tilde{W}}c^2+C^6_{\Phi \tilde{B}}s^2
+C^6_{\Phi \tilde{W}B}cs),
\label{a1a2a3}
\eea
where $c$ and $s$ stand for the cosine and sine of the weak mixing angle respectively,
$a_1,a_2,a_3$ are dimensionless complex numbers and $v = 246$~GeV is the electroweak vacuum expectation value.
Notice that the signs before $a_2$ and $a_3$ are same as in \cite{Gao:2010qx,Khachatryan:2014kca,Sirunyan:2019twz}, but have an additional minus sign from the definition in \cite{Chen:2013ejz}.
$Z_{\mu}$ is $Z$ boson field, $Z_{\mu\nu}=\partial_{\mu} Z_{\nu}-\partial_{\nu} Z_{\mu}$ is the field strength tensor of the $Z$ boson and  $\tilde{Z}_{\mu\nu}=\frac{1}{2}\epsilon_{\mu\nu\rho\sigma}Z^
{\rho\sigma}$ represents its dual field strength.
The loop corrections in SM can contribute similarly as 
the $a_2$ and $a_3$ terms. 
Quantitatively, the one-loop correction can contribute to $a_2$ term with small contributions
$\mathcal{O}(10^{-2}-10^{-3})$,
while the $a_3$ term appear in SM only at a three-loop level and thus has
a even smaller contribution~\cite{Khachatryan:2014kca}. Therefore, only if the contributions from the $a_2$ and $a_3$ terms are larger than these loop contributions can we consider them as from new physics.
 
The $HZZ$ interaction vertex from Eq.~\eqref{lagrangian} is
\beq
\Gamma_a^{\mu\nu}(k,k^{\prime})=i\frac{2}{v}\sum_{i=1}^3a_i\Gamma_{a,i}^{\mu\nu}(k,k^{\prime})=i\frac{2}{v}[a_1M_Z^2g^{\mu\nu}-2a_2(k^{\nu} k^{\prime\mu}-k\cdot k^{\prime}g^{\mu\nu})-2a_3\epsilon^{\mu\nu\rho\sigma} k_{\rho}k^{\prime}_{\sigma} ]~,
\label{eqn:gamma}
\eeq
where $k$,$k^{\prime}$ are the momenta of the two $Z$ bosons.
It is worthy to notice that the $HZZ$ vertices in the SM are
\beq
\Gamma_\text{SM}^{\mu\nu}(k,k^{\prime})=i\frac{2}{v}M_Z^2g^{\mu\nu}~,
\eeq
so the Lorentz structure of the $a_1$ term is same as the SM case.
While the $a_2$ and $a_3$ terms have different Lorentz structures, which 
represent non-SM $CP$-even and $CP$-odd cases respectively. 

\subsection{Helicity amplitude of the process $gg\rightarrow H\rightarrow ZZ\to2e2\mu$ }
\begin{figure}[htbp]
\centerline{ \includegraphics[width=0.5\textwidth]{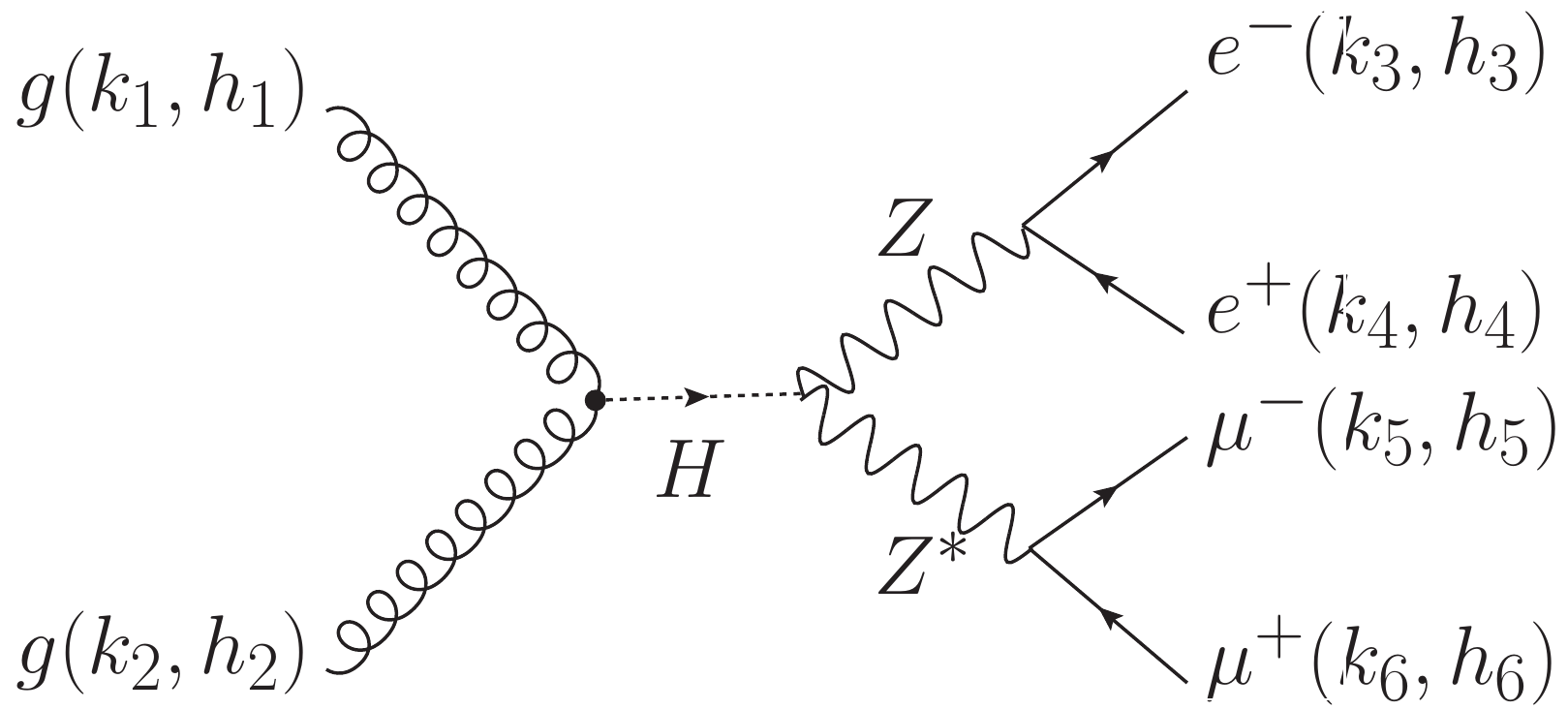} }
\captionsetup{singlelinecheck = false, justification=justified}
\caption{ Feynman diagram of the Higgs-mediated process $gg \rightarrow H \rightarrow ZZ\to2e2\mu$. The black dot represents an effective $ggH$ coupling from loop contributions.}
\label{ggHZZ}
\end{figure}

The total helicity amplitude for the process $gg\rightarrow H\rightarrow ZZ\to2e2\mu$ in Fig.~{\ref{ggHZZ}} is composed of three individual amplitudes $A^H_{\rm SM}, A^H_{CP-\rm{even}}$ and $A^H_{CP-\rm{odd}}$,
which have the same production process but different Higgs decay modes according to the three kinds of $HZZ$ vertices in Eq.~\eqref{eqn:gamma}. The specific formulas are
\bea
&&~~~~\mathcal{A}^{gg\to H\rightarrow ZZ\to2e2\mu}
(1_g^{h_1},2_g^{h_2},3_{e^-}^{h_3},4^{h_4}_{e^+},5_{\mu^-}^{h_5},6^{h_6}_{\mu^+})\\ \label{eqn:a3}
&&=[a_1\mathcal{A}^{H}_{\rm SM}+a_2\mathcal{A}^{H}_{CP-\rm{even}}+a_3\mathcal{A}^{H}_{CP-\rm{odd}}]
(1_g^{h_1},2_g^{h_2},3_{e^-}^{h_3},4^{h_4}_{e^+},5_{\mu^-}^{h_5},6^{h_6}_{\mu^+})~, \\ \label{eqn:aphd}
&&=\mathcal{A}^{gg\rightarrow H}(1_g^{h_1},2_g^{h_2})\times
\frac{P_H(s_{12})}{s_{12}}\times \sum_{i=1}^3 a_i\mathcal{A}_i^{H\rightarrow ZZ\to2e2\mu}(3_{e^-}^{h_3},4^{h_4}_{e^+},5_{\mu^-}^{h_5},
6^{h_6}_{\mu^+})~,
\eea
where $h_i$ $(i=1\cdots 6)$ are helicity indices of external particles, $s_{ij}=(k_i+k_j)^2$ and $P_H(s)=\frac{s}{s-M_H^2+iM_H\Gamma_H}$ is the Higgs propagator.

The production part $\mathcal{A}^{gg\to H}(1^{h_1}_g,2^{h_2}_g)$ is the helicity amplitude of gluon-gluon
fusion to Higgs process, in which $h_1, h_2$ represent the helicities of gluons with outgoing momenta.
For all the other helicity amplitudes in this paper, we also keep the convention that the momentum of each external particle is outgoing.
When writing the helicity amplitudes, we adopt the conventions used in~\cite{Dixon:1996wi, Campbell:2013una}:
\bea
&&\langle ij \rangle = \bar{u}_-(p_i) u_+(p_j), \qquad  ~~{[ ij ]} = \bar{u}_+(p_i) u_-(p_j)~,\nonumber\\
&&\langle ij \rangle[ ji ] = 2 p_i \cdot p_j,  \qquad ~~s_{ij} = (p_i+p_j)^2,
\eea
and we have
 \bea
 \mathcal{A}^{gg\to H}(1^{+}_g,2^{+}_g)&=&\frac{2c_g}{v}[12]^2~, \nonumber\\
 \mathcal{A}^{gg\to H}(1^{-}_g,2^{-}_g)&=&\frac{2c_g}{v}\langle12\rangle^2~.
 \label{eqn:ggh}
\eea
To keep the $ggH$ coupling consistent with SM, we make
\beq
\frac{c_g}{v}=\frac{1}{2}\sum_f\frac{\delta^{a b}}{2}\frac{i}{16\pi^2}g^2_s4e
\frac{m_f^2}{2M_W s_W}\frac{1}{s_{12}}[2+s_{12}(1-\tau_H)C^{\gamma\gamma}_0(m_f^2)]~,
\label{eqn::FggH}
\eeq
with 
\beq
C^{\gamma \gamma}_0(m^2) =  2\tau_H f(\tau_H)/4m^2~, \tau_H=4m^2/M^2_{H},
\eeq
\beq
f(\tau) = \left\{ \begin{array}{ll}
{\rm arcsin}^2 \sqrt{1/\tau} & \tau \geq 1 \\
-\frac{1}{4} \left[ \log \frac{1 + \sqrt{1-\tau } }
{1 - \sqrt{1-\tau} } - i \pi \right]^2 \ \ \ & \tau <1~
\end{array} \right.~,
\label{eqn:c0gammagamma}
\eeq
where $a, b=1,...,8$ are $SU(3)_c$ adjoint representation indices for the gluons,
the index $f$ represents quark flavor and $C^{\gamma\gamma}_{0}(m^2)$ is the Passarino-Veltman three-point scalar function \cite{Passarino:1978jh,Chen:2017plj}. 

The decay part $\mathcal{A}^{H\rightarrow ZZ\to2e2\mu}(3_{e^-}^{h_3},4^{h_4}_{e^+},5_{\mu^-}^{h_5},6^{h_6}_{\mu^+})$ is the helicity amplitude of
the process $H\rightarrow ZZ\to e^-e^+\mu^-\mu^+$, which have three sources according to the three types of vertices as written in Eq.~\eqref{eqn:gamma}.
Correspondingly we write it as
\beq
\mathcal{A}^{H\rightarrow ZZ\to2e2\mu}(3_{e^-}^{h_3},4^{h_4}_{e^+},5_{\mu^-}^{h_5},6^{h_6}_{\mu^+})=\sum_{i=1}^3 a_i\mathcal{A}_i^{H\rightarrow ZZ\to2e2\mu}(3_{e^-}^{h_3},4^{h_4}_{e^+},5_{\mu^-}^{h_5},6^{h_6}_{\mu^+})
\eeq
with
\bea
\label{eqn:amp1}
&& \mathcal{A}_1^{H\rightarrow ZZ\to2e2\mu}(3^-_{e^-},4^+_{e^+},5^-_{\mu^-},6^+_{\mu^+})=
f \times  l^2_e\frac{M_W^2}{\cos^2\theta_W}\langle35\rangle[46] , \\ \nonumber \label{eqn:amp2}
&& \mathcal{A}_2^{H\rightarrow ZZ\to2e2\mu}(3^-_{e^-},4^+_{e^+},5^-_{\mu^-},6^+_{\mu^+})=
f\times l^2_e\times \\
&&
\Big[2k\cdot k^{\prime}\langle35\rangle[46]
 +\big(\langle35\rangle[45]+
\langle36\rangle[46]\big)\big(\langle35\rangle[36]+\langle45\rangle[46]\big) \Big],  \\ \nonumber
&& \mathcal{A}_3^{H\rightarrow ZZ\to2e2\mu}(3^-_{e^-},4^+_{e^+},5^-_{\mu^-},6^+_{\mu^+})=
f\times l^2_e\times (-i) \times  \\ \nonumber
&& \Big[2\big(k\cdot k^{\prime}+\langle46\rangle[46]\big)\langle35\rangle[46]
+\langle35\rangle[45]\big(\langle35\rangle[36]+\langle45\rangle[46]\big)      \\
&&+\langle36\rangle[46]\big(\langle35\rangle[36]-\langle45\rangle[46]\big)\Big]~\label{eqn:amp3}.
\eea
and
\beq
f=-2ie^3\frac{1}{M_W \sin\theta_W}\frac{P_Z(s_{34})}{s_{34}}\frac{P_Z(s_{56})}{s_{56}}~,
\eeq
where $P_Z(s)=\frac{s}{s-M_Z^2+iM_Z\Gamma_Z}$ is the $Z$ boson propagator, $M_Z$,$M_W$ are the masses of the $Z$,$W$ bosons, $\theta_W$ is the Weinberg angle, $l_e$ and $r_e$ ( will appear for other helicity combinations) are the coupling factors of the $Z$ boson to left-handed and right-handed leptons:
\beq
l_e=\frac{-1+2\sin^2\theta_W}{\sin(2\theta_W)}~, r_e=\frac{2\sin^2\theta_W}{\sin(2\theta_W)}~.
\eeq
In Eq.s~\eqref{eqn:amp1}\eqref{eqn:amp2}\eqref{eqn:amp3}, we only show the case in which
the helicities of the four leptons
($h_3,h_4,h_5,h_6$) are equal to ($-,+,-,+$). As for the other three non-zero helicity combinations
 ($-,+,+,-$), ($+,-,-,+$), ($+,-,+,-$), their helicity amplitudes are similar to Eq.s~\eqref{eqn:amp1}\eqref{eqn:amp2}\eqref{eqn:amp3}, but with some exchanges such as
\beq
l_e \leftrightarrow r_e ~,~ 4 \leftrightarrow 6~,~ 3 \leftrightarrow 5~,~ []\leftrightarrow \langle\rangle~.
\eeq
Their specific formulas are shown in Appendix~\ref{app:amp}.

\subsection{Helicity amplitude of the box process $gg\rightarrow ZZ\to2e2\mu$}
\begin{figure}[htbp]
\centering
\centerline{ \includegraphics[width=0.5\textwidth]{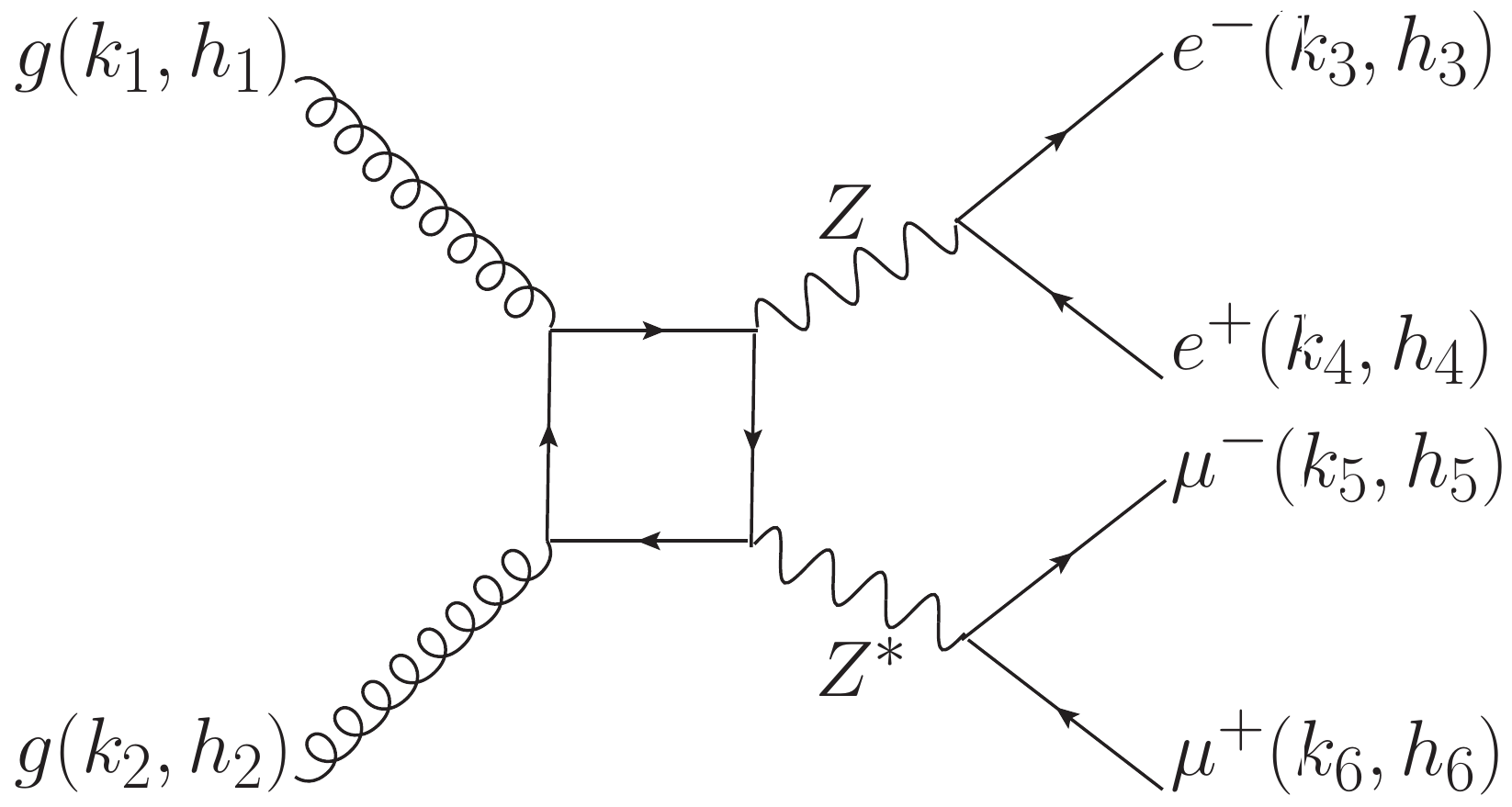} }
\caption{ Feynman diagram of the box process $gg \rightarrow ZZ\to2e2\mu$~.}
\label{image3}
\end{figure}

The box process $gg\rightarrow ZZ\to2e2\mu$ is a continuum background of the Higgs-mediated $gg\to H\to2e2\mu$ process. The interference between these two kinds of processes can have nonnegligible contribution in the off-shell Higgs region.
The Feynman diagram of the process $gg\rightarrow ZZ\to2e2\mu$ is a box diagram
which is induced by fermion loops (see Fig.~\ref{image3}).
The helicity amplitude $A^{gg\rightarrow ZZ\to2e2\mu}_\text{box}$
has been calculated analytically
and coded in \texttt{MCFM8.0} package.
Another similar calculation that using a different method can be found in
\texttt{gg2VV} code~\cite{Binoth:2008pr}.

\subsection{Helicity amplitude of the process $gg\rightarrow H\rightarrow ZZ \to 4\ell$}

\begin{figure}[htbp]
\centering
\begin{minipage}[c]{0.5\textwidth}
\centering
\includegraphics[width=0.9\textwidth]{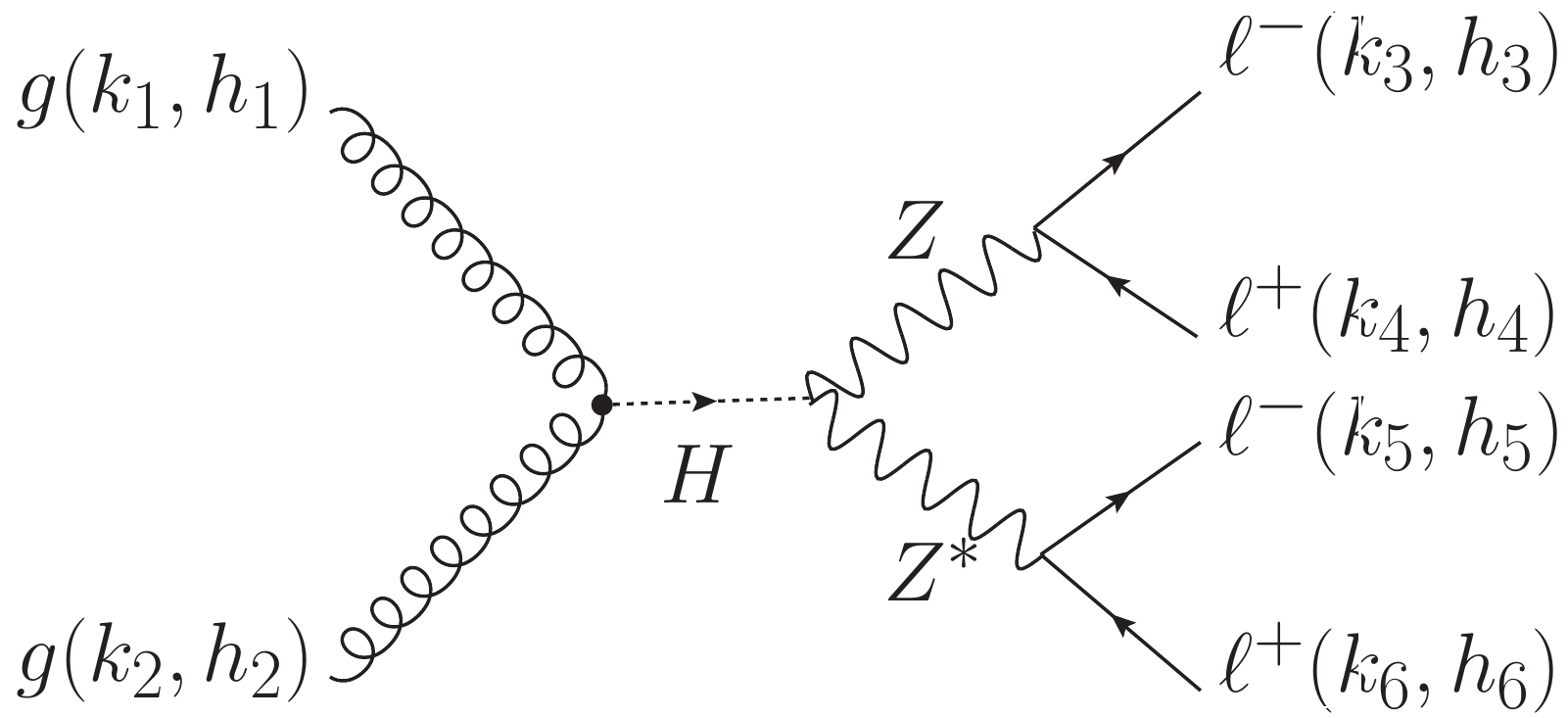}\\
(a)
\end{minipage}%
\begin{minipage}[c]{0.5\textwidth}
\centering
\includegraphics[width=0.9\textwidth]{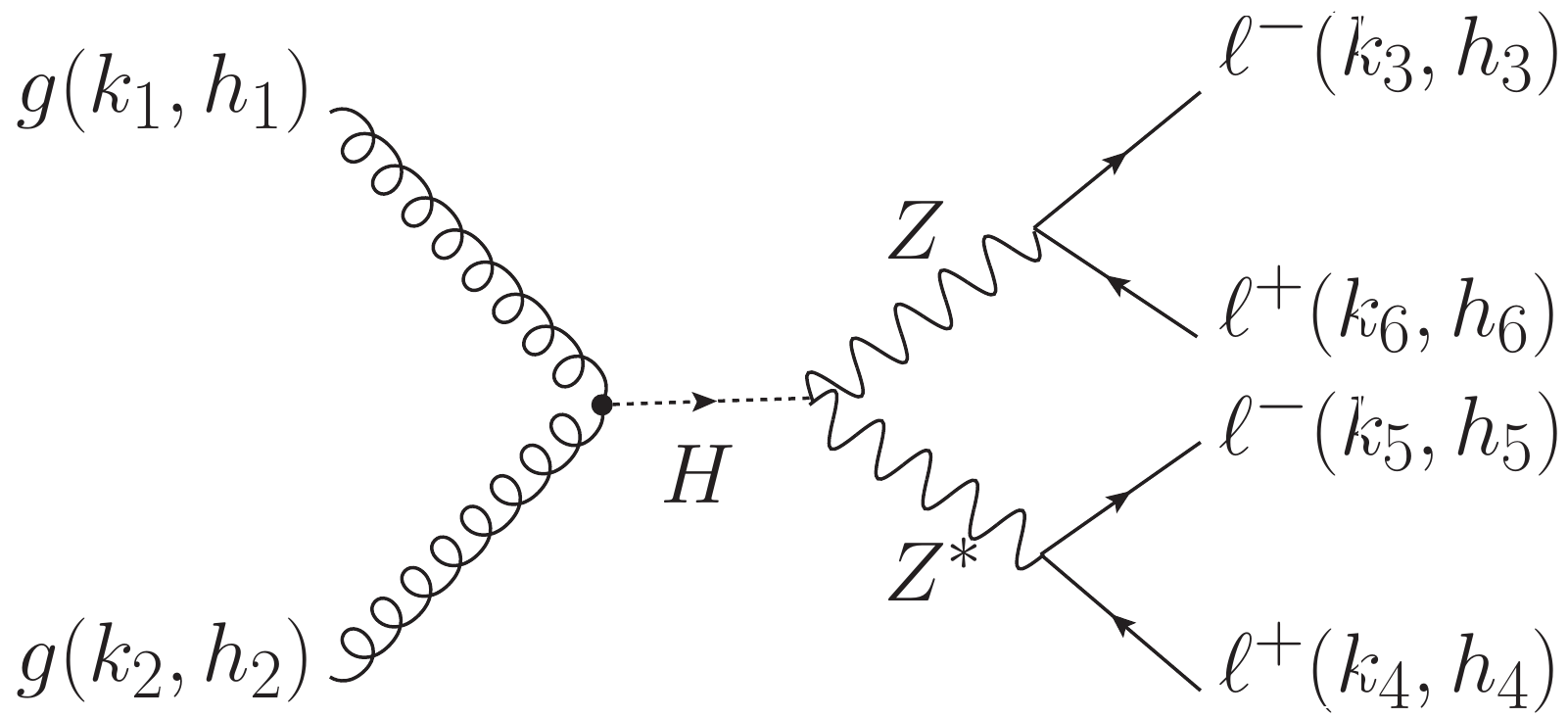}\\
(b)
\end{minipage}
\caption{ The Feynman diagrams of the process $gg \rightarrow H\rightarrow  ZZ\to 4\ell$, where $4\ell=4e\ \mbox{or}\ 4\mu$. Note that diagram (b)
is obtained by swapping the two positive charged leptons (4$\leftrightarrow$6) in diagram (a). }
\label{image4}
\end{figure}

The process $gg\rightarrow H\rightarrow ZZ \to 4\ell$ with identical $4e$ or $4\mu$ final states can also be used to probe the anomalous $HZZ$ couplings.
In SM the differential cross sections of the $4\ell$ (include both $4e$ and $4\mu$ ) and $2e2\mu$ processes are nearly the same in both on-shell and off-shell Higgs  regions~\cite{Ellis:2014yca},
which indicates adding the $4e/4\mu$ process can almost double experimental statistics.
This situation can probably be similar for the anomalous Higgs-mediated processes.
The $4e/4\mu$ Feynman diagrams consist of two different topology structures as shown in Fig.~\ref{image4}.
Fig.~\ref{image4}(b) is different from Fig.~\ref{image4}(a) just by swapping the positive charged leptons (4$\leftrightarrow$6). The helicity amplitude of each diagram is similar to the former $2e2\mu$ cases but need to be multiplied by a symmetry factor $\frac{1}{2}$.
 While calculating the total cross section the interference term between Fig.~\ref{image4}(a) and (b)
 need an extra factor of -1 comparing to the self-conjugated terms
 because it connects all of the decayed leptons in one fermion loop while each self-conjugated term
has two fermion loops.
After considering these details, the summed cross section of $4e$ and $4\mu$ processes
is comparable to the $2e2\mu$ process. More details are shown in the following numerical results.

\section{Numerical result\label{section:numerical result}}

In this section we present the integrated cross sections and differential distributions in both
on-shell and off-shell Higgs regions, especially
 the interference between anomalous Higgs-mediated processes and SM processes.

\subsection{The cross sections}

To compare theoretical calculation with experimental observation at LHC,
we need to further calculate the cross sections at hadron level.
From helicity amplitude to the cross section,
there need two more steps. Firstly we should
sum and square the amplitudes to
get the differential cross section at
parton level, then integrate phase space and parton distribution function (PDF)
to get the cross section at hadron level.
As following we show these two steps conceptually.

The squared amplitude in the differential cross section at parton level $d\hat{\sigma}(s_{12})$ is
\bea
& & \left|\mathcal{A}_\text{box}^{gg\to ZZ\to 4\ell}+\mathcal{A}^{gg\to H \to ZZ\to 4\ell}\right|^2  \\
&= & \left|\mathcal{A}_\text{box}^{gg\to ZZ\to 4\ell}+\mathcal{A}^{H}_{\rm SM}+
a_1\mathcal{A}^{H}_{\rm SM}+a_2\mathcal{A}^{H}_{CP-\rm{even}}+a_3\mathcal{A}^{H}_{CP-\rm{odd}}
\right|^2~.
\label{eqn:dsigma}
\eea
After expanding it, there left self-conjugated terms
and interference terms that have different amplitude sources.
As in the next step the integral of phase space and PDF are same for each term,
we note the integrated cross sections separately by the amplitude sources, which are

\beq
 \sigma_{k,l}\sim
\left\{
\begin{array}{ll}
|\mathcal{A}_k|^2, \quad\quad & k=l ;\\
2\text{Re}(\mathcal{A}^\ast_k\mathcal{A}_l), \quad\quad& k\ne l ,\\
\end{array}  \right.
\label{eq:sigmabox}
\eeq
where $k,l=$~\{box, SM, $CP$-even, $CP$-odd\}.
The superscripts of $\mathcal{A}$ are omitted for brevity.

\subsection{Numerical results for $gg\to 2e 2\mu$ process}

We make the integral of phase space and the PDF in the \texttt{MCFM}~8.0 package~\cite{Campbell:2015qma,Boughezal:2016wmq}.
The simulation is performed for the proton-proton collision at the
center-of-mass energy $\sqrt{s}=13$~TeV.
The Higgs mass is set to be $M_H=125 \rm~GeV$.
The renormalization $\mu_r$ and factorization scale $\mu_f$ are
set as the dynamic scale $m_{4\ell}/2$.
For PDF we choose the leading-order MSTW 2008 PDFs MSTW08LO~\cite{Martin:2009iq}.
Some basic phase space cuts are exerted as follows, which are similar to
the event selection cuts used in CMS experiment~\cite{CMS-PAS-HIG-13-002}.
\begin{equation}
\begin{aligned}
&P_{T,\mu}>5\rm~GeV, \ |\eta_{\mu}|<2.4~,\\
&P_{T,e}>7\rm~GeV, \ |\eta_{e}|<2.5~,\\
&\ m_{\ell\ell}>4 \rm~GeV, \ m_{4\ell}>100\rm~GeV~.\\
\end{aligned}
\end{equation}
Besides, for the $2e2\mu$ channel,
the hardest (second-hardest) lepton should satisfy $P_T> 20~(10)\rm~GeV$;
one pair of leptons with the
same flavour and opposite charge is required to have $40 \rm~GeV<m_{\ell^+\ell^-}<120 \rm~GeV$ and the other pair needs to fulfill $12 \rm~GeV <m_{\ell^+\ell^-}<120 \rm~GeV$.
For the $4e$ or $4\mu$ channel, four oppositely charge lepton pairs exist as $Z$ boson candidates. The
selection strategy is to first choose one pair nearest to the $Z$ boson mass as one $Z$ boson,
then consider the left two leptons as the other $Z$ boson. The other requirements are similar to the
$2e2\mu$ channel.

\begin{table*}[!htp]
\begin{floatrow}

\begin{minipage}{0.5\linewidth}
\centering

\begin{tabular}{|c|c|c|c|c|c|}
 \hline
 \multicolumn{6}{|c|}{13~\text{TeV}, $m_{2e2\mu}<130\rm~GeV$, on-shell} \\
 \hline
 \multicolumn{2}{|c|}{\multirow{2}{*}{$\sigma_{k,l}$(fb)}} & \multirow{2}{*}{box} &
\multicolumn{3}{c|}{\footnotesize Higgs-med.}\\
 \cline{4-6}
 \multicolumn{2}{|c|}{} & & {\footnotesize SM} & {\footnotesize $CP$-even} & {\footnotesize $CP$-odd}  \\
\hline
\multicolumn{2}{|c|}{box} & 0.024 & 0 & 0 & 0 \\
\hline
\multirow{3}{*}{ \begin{tabular}{c} \rotatebox{90}{\footnotesize Higgs-med.} \end{tabular}} &
 {\footnotesize SM} & 0 & 0.503 & 0.558 & 0\\
\cline{2-6}
 & \footnotesize $CP$-even & 0 & 0.558 & 0.202 & 0 \\
\cline{2-6}
 & \footnotesize $CP$-odd & 0 & 0 & 0 & 0.075 \\
\hline
 \end{tabular}

\end{minipage}

\hfill

\begin{minipage}{0.5\linewidth}

\begin{tabular}{|c|c|c|c|c|c|}
 \hline
 \multicolumn{6}{|c|}{13~\text{TeV}, $m_{2e2\mu}>220\rm~GeV$, off-shell} \\
 \hline
 \multicolumn{2}{|c|}{\multirow{2}{*}{$\sigma_{k,l}$(fb)}} & \multirow{2}{*}{box} &
\multicolumn{3}{c|}{\footnotesize Higgs-med.}\\
 \cline{4-6}
 \multicolumn{2}{|c|}{} & & {\footnotesize SM} & {\footnotesize $CP$-even} & {\footnotesize $CP$-odd}  \\
\hline
\multicolumn{2}{|c|}{box} & 1.283 & -0.174 & -0.571 & 0 \\
\hline
\multirow{3}{*}{ \begin{tabular}{c} \rotatebox{90}{\footnotesize Higgs-med.} \end{tabular}} &
 {\footnotesize SM} & -0.174 & 0.100 & 0.137 & 0\\
\cline{2-6}
 & \footnotesize $CP$-even & -0.571 & 0.137 &  0.720 & 0 \\
\cline{2-6}
 & \footnotesize $CP$-odd & 0 & 0 & 0 & 0.716 \\
\hline
 \end{tabular}

\end{minipage}

\caption{ The cross sections of $gg\to2e2\mu$ processes in proton-proton collision at
center-of-mass energy $\sqrt{s}=13$~TeV with $a_1=0, a_2=a_3=1$ in Eq.~\eqref{eqn:gamma}.
}
\label{table:13tevsigma}
\end{floatrow}
\end{table*}

Table~\ref{table:13tevsigma} shows the cross sections $\sigma_{k,l}$ with
$k,l=$~\{box, SM, $CP$-even, $CP$-odd\}
while $a_1,a_2,a_3$ are all set to 1 for convinience. The cross section values
can be converted easily by multiplying a scale factor for small $a_i$s.
 In the left and right panels, the integral regions
of $m_{4\ell}$ are separately set as $m_{4\ell}<130~\text{GeV}$ and $m_{4\ell}>220~\text{GeV}$,
which correspond to the on-shell and off-shell Higgs regions, respectively.
Next we focus on two kinds of interference effects: the interference
between each Higgs-mediated process and box continuum background, denoted as $\sigma_{\text{box},l}$ (or $\sigma_{l,\text{box}}$) with $l\ne\text{box}$; and the interference between different Higgs-mediated processes, denoted as $\sigma_{k,l}$ with $k,l\ne {\rm box}$.

The interference terms
between Higgs-mediated processes
and the continuum background $\sigma_{\text{box},l}$ are all zeros in on-shell Higgs region, but relatively sizeble in the off-shell regions
except for the cases with the $CP$-odd Higgs-mediated process as shown in
Table~\ref{table:13tevsigma}.
There is an interesting reason for it. As from Eq.~\eqref{eqn:a3}\eqref{eqn:aphd}\eqref{eq:sigmabox},
\bea
\nonumber
\sigma_{\text{box},l}&\sim & 2\text{Re}(\mathcal{A}^\ast_\text{box}\mathcal{A}_l)~, \\
\nonumber
&\sim & 2\text{Re}\big(\mathcal{A}^\ast_{\rm box} \mathcal{A}^{gg\to H}P_H(s_{12})\mathcal{A}_i\big)~,\\
&\sim & 2\frac{(s_{12}-M^2_H)\text{Re}\big(\mathcal{A}^\ast_{\rm box}\mathcal{A}^{gg\to H}\mathcal{A}_i\big)+M_H\Gamma_H\text{Im}\big(\mathcal{A}^\ast_{\rm box}\mathcal{A}^{gg\to H}\mathcal{A}_i\big)}
{(s_{12}-M^2_H)^2+M^2_H\Gamma^2_H}~,
\label{eqn:sigmaint}
\eea
which means the integrand of $\sigma_{\text{box},l}$ consists of two parts, one is antisymmetric
around $M^2_H$, the other is proportional to $M_H\Gamma_H\text{Im}\big(\mathcal{A}^\ast_{\rm box}\mathcal{A}^{gg\to H}\mathcal{A}_i\big)$.
The first part can be largely suppressed almost to zero
 in the integral with an integral region symmetric around $M_H$.
The second part is also suppressed not only by the small factor
of $\Gamma_H/M_H$
but also by a small value of $\text{Im}\big(\mathcal{A}^\ast_{\rm box}\mathcal{A}^{gg\to H}\mathcal{A}_i\big)$ in the on-shell Higgs region.
By contrary, in the off-shell Higgs region
 the integral regions are not symmetric around $M_H$ but in one side larger than $M_H$, which makes the first term have some non-zero contribution.
Both the first and the second terms can also be enhanced
when $\sqrt{s_{12}}$ is a little larger than twice of the top quark mass.
That is because the $gg\to H$ process is induced mainly by
top quark loop,
both the real part and the imaginary part of the amplitude
(Re$\mathcal{A}^{gg\to H}$ and Im$\mathcal{A}^{gg\to H}$)
can be enhanced when $\sqrt{s_{12}}$ is just larger than the $2M_t$ threshold (see Eq.~\eqref{eqn::FggH}).
Then $\text{Im}\big(\mathcal{A}^\ast_{\rm box}\mathcal{A}^{gg\to H}\mathcal{A}_i\big)$ can have a larger value, even though
the relative contribution from the second term can be still suppressed by the smallness of the factor $\Gamma_H/M_H$.
In conclusion, mainly due to the nonsymmetric integral region
 and some enhancement of $\mathcal{A}^{gg\to H}$,
the interferece contribution
in the off-shell Higgs region becomes comparable with the self-conjugated
contributions.

It is also worthwhile to point out there is no cross section contribution from the
interference between the $CP$-odd Higgs-mediated process and other three processes, which include the continuumm background process, SM Higgs-mediated process and anomalous $CP$-even Higgs-mediated process. It is because there is an antisymmetric tensor $\epsilon^{\mu\nu\rho\sigma}$ in the $CP$-odd $HZZ$ interaction vertex (see last term in Eq.~\eqref{eqn:gamma}), while in the other three processes, the two indices are symmetrically paired
 and so the contract of the indices makes the interference term zero.
Nevertheless, these $CP$-odd interference term
 can show angular distributions, include polar angle
distribution of $\ell$ in $Z$ boson rest frame and azimuthal angular distribution
between two $z$ decay planes~\cite{Buchalla:2013mpa,Beneke:2014sba}, even though its contribution to the total cross scetion is still zero.

The interference between $CP$-even Higgs-mediated process
and SM Higgs-mediated process
is nonnegligible both in on-shell and off-shell Higgs regions.
In on-shell Higgs region, the contribution from interfernce terms
is larger than that from the self-conjugated terms. Furthermore,
for $a_1=0, a_2=-1$
choice(as in \cite{Chen:2013ejz}), the interference terms would have a minus sign, comparing to
the relative values in Table~\ref{table:13tevsigma}, which makes the total contribution of
$CP$-even Higgs-mediated process beyond SM a destructive effect.
In the off-shell region, the $CP$-even Higgs-mediated process
have two interference terms, separately between SM Higgs-mediated process
and the box process. These two interference terms have opposite sign,
which means they cancel each other partly. Even though,
the summed interfernce effect is still comparable to the self-conjugated contribution.

\begin{figure}[!htbp]
\centering
\centerline{\includegraphics[width=0.8\textwidth]{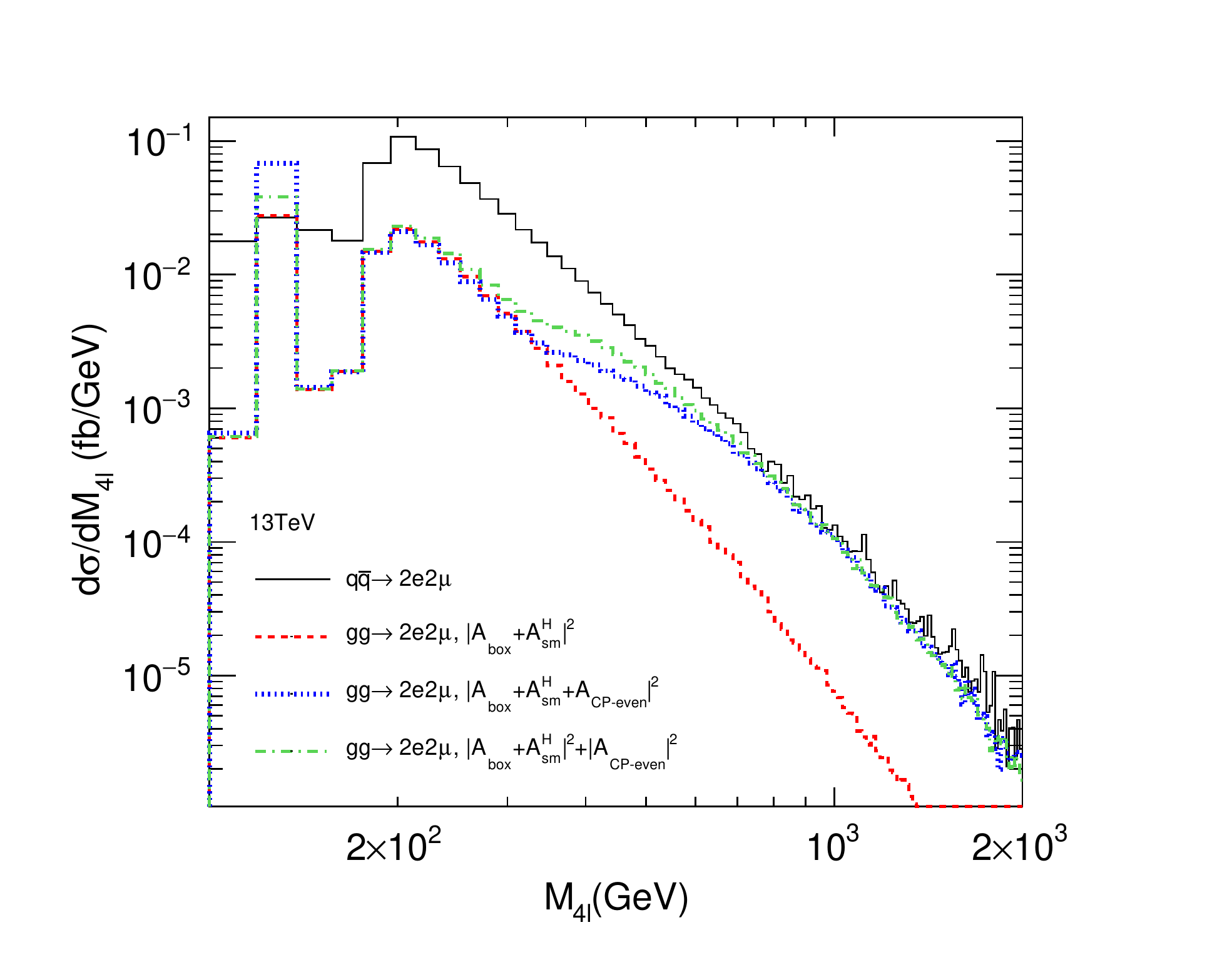} }
\caption{ Differential cross sections of the $gg\to2e2\mu$ processes and $q\bar{q}\to2e2\mu$ process in proton-proton collision at $\sqrt{s}=13$~TeV  with $a_2=1, a_1=a_3=0$ in Eq.~\eqref{eqn:gamma}.
}
\label{fig:dsigma_int}
\end{figure}

Fig.~\ref{fig:dsigma_int} shows the differential cross sections.
The black histogram is from its main background process $q\bar{q}\to 2e 2\mu$,
 which is a huge background but still controllable. The red dashed histogram
 is from the SM $gg\to 2e 2\mu$
processes including contributions from both
the box and SM Higgs-mediated amplitudes.
The blue dotted histogram adds contribution from the $CP$-even Higgs-mediated amplitude to
the SM signal and background amplitudes. Therefore three kinds of interference terms are included.
For comparison,
we also show the green dashed-dotted histogram without interference terms
 from  $CP$-even Higgs amplitudes with others,
so the interference contribution can be calculated by the difference
between blue and green histograms.
In the on-shell region we can see the $CP$-even Higgs-mediated
process have a total positive contribution (blue histogram) compare to the SM process(red histogram), while the green histogram shows
the main positive contribution is from the interference term.
In the off-shell region, the interference contribution
is obvious in $200~\rm GeV < m_{4\ell} < 600~\rm GeV$ region.
There is a bump in blue and green histograms when $m_{4\ell}\approx350$~GeV,
 which is caused by the total cross section of the $CP$-even Higgs-mediated
process increase suddenly
beyond the $2M_t$ (twice of the top quark mass) threshold.
The differential cross section for the $CP$-odd Higgs-mediated
process is similar to the green histogram in off-shell region since it has no interference
 contribution after the angular distributions being integrated.

The numerical results at center-of-mass energy $\sqrt{s}=8$~TeV
are shown in Table~\ref{table:8tevsigma} in Appendix \ref{table8tev}. By comparing them to
the results at $\sqrt{s}=13$~TeV in Table~\ref{table:13tevsigma},
we can find that
each cross section is decreased by about one or two times
and their relative ratios have some minor changes. That can be caused by
both PDF functions and kinematic distributions.

\subsection{Numerical results for $gg\to 4e/4\mu$ processes}

\begin{table*}[!htp]
\begin{floatrow}

\begin{minipage}{0.5\linewidth}
\centering

\begin{tabular}{|c|c|c|c|c|c|}
 \hline
 \multicolumn{6}{|c|}{13~\text{TeV},~$m_{4e/4\mu}<130\rm~GeV$, on-shell} \\
 \hline
 \multicolumn{2}{|c|}{\multirow{2}{*}{$\sigma_{k,l}$(fb)}} & \multirow{2}{*}{box} &
\multicolumn{3}{c|}{\footnotesize Higgs-med.}\\
 \cline{4-6}
 \multicolumn{2}{|c|}{} & & {\footnotesize SM} & {\footnotesize $CP$-even} & {\footnotesize $CP$-odd}  \\
\hline
\multicolumn{2}{|c|}{box} & 0.045 & 0 & 0 & 0 \\
\hline
\multirow{3}{*}{ \begin{tabular}{c} \rotatebox{90}{\footnotesize Higgs-med.} \end{tabular}} &
 {\footnotesize SM} & 0 & 0.540 & 0.568 & 0 \\
\cline{2-6}
 & \footnotesize $CP$-even & 0 & 0.568 & 0.186 & 0 \\
\cline{2-6}
 & \footnotesize $CP$-odd & 0 & 0  & 0 & 0.060  \\
\hline
 \end{tabular}

\end{minipage}

\hfill

\begin{minipage}{0.5\linewidth}

\begin{tabular}{|c|c|c|c|c|c|}
 \hline
 \multicolumn{6}{|c|}{13~\text{TeV},~$m_{4e/4\mu}>220\rm~GeV$, off-shell} \\
 \hline
 \multicolumn{2}{|c|}{\multirow{2}{*}{$\sigma_{k,l}$(fb)}} & \multirow{2}{*}{box} &
\multicolumn{3}{c|}{\footnotesize Higgs-med.}\\
 \cline{4-6}
 \multicolumn{2}{|c|}{} & & {\footnotesize SM} & {\footnotesize $CP$-even} & {\footnotesize $CP$-odd}  \\
\hline
\multicolumn{2}{|c|}{box} & 1.303 & -0.176  & -0.575 & 0 \\
\hline
\multirow{3}{*}{ \begin{tabular}{c} \rotatebox{90}{\footnotesize Higgs-med.} \end{tabular}} &
 {\footnotesize SM}  & -0.176 & 0.101 & 0.137 & 0 \\
\cline{2-6}
 & \footnotesize $CP$-even & -0.575 & 0.137 &  0.740 & 0  \\
\cline{2-6}
 & \footnotesize $CP$-odd & 0 & 0 & 0 & 0.708 \\
\hline
\end{tabular}

\end{minipage}

\caption{ Cross sections of $gg\to4e/4\mu$ processes in proton-proton collisions at
center-of-mass energy $\sqrt{s}=13$~TeV with $a_1=0, a_2=a_3=1$ in Eq.~\eqref{eqn:gamma}.
}
\label{table:13tevsigma_4l}
\end{floatrow}
\end{table*}

The cross sections of $gg\to4e/4\mu$ processes are listed in
Table~\ref{table:13tevsigma_4l} (Table~\ref{table:8tevsigma_4l} in Appendix~\ref{table8tev})
 for comparison and next use.
Here $gg\to4e/4\mu$ represents the sum of $gg\to4e$ and $gg\to 4\mu$.
Comparing Table~\ref{table:13tevsigma_4l} with
Table~\ref{table:13tevsigma},
the numbers in the right panels are similar, while
the numbers in the left panels have relatively large differences.
That is mainly because the different selection cuts~\cite{Ellis:2014yca}.
If apply the $4e/4\mu$ selection cuts to the $gg\to 2e2\mu$ process,
$\sigma_{\rm box, box}$ in the left panels can become similar.

\section{ Constraints: a naive estimation \label{section:constraints}}

In this section we show a naive estimation to constrain $a_1$, $a_2$ and $a_3$
by using the data in both the on-shell and off-shell Higgs regions.

First, we estimate the expected number of events $N^{\rm exp}(a_1, a_2,a_3)$
in the off-shell Higgs region, which is defined as the contribution from the processes with anomalous couplings after excluding the pure SM contributions.

A theoretical observed total number of events should be
\beq
N^{\rm theo}(a_1,a_2,a_3)= \sigma_{\rm tot}\times \mathcal{L}\times
k\times\epsilon~,
\eeq
where $\sigma_{\rm tot}$ is the total cross section, 
$\mathcal{L}$ is the integrated luminosity, 
$k$ represents the $k$-factor and $\epsilon$ is the total  efficiency.

The simulation in the CMS experiment~\cite{Sirunyan:2019twz}
with an integrated luminosity of
$\mathcal{L}\sim 80$~fb$^{-1}$ at $\sqrt{s}=13$~TeV
shows that
for the $gg\to4\ell$ process, the expected
numbers of events in the off-shell Higgs region ($m_{4\ell}>220\rm~GeV$) can be divided into two categories:
$N_{gg~~ {\rm signal}}=20.3$ and $N_{gg~~{\rm interference}}=-34.4$,
where the subscript ``$gg$ signal'' represents the SM Higgs-mediated signal term,
``$gg$ interference'' represents the interference term between
SM Higgs-mediated process and the box process.
For high-order corrections that may change the $k$-factor, some existing studies~\cite{Caola:2015psa,Melnikov:2015laa,Campbell:2016ivq,Caola:2016trd} show that the loop corrections on the box diagram~(Fig.\ref{image3}) and the Higgs-mediated diagram are different. Therefore, we also group the expected event number contributed from the anomalous couplings into two categories.
\bea
\nonumber
&&N^{\rm exp}(a_1,a_2,a_3)\\
\nonumber
&=&
\frac{N_{gg~~{\rm signal}}}{\sigma^{H}_{\text{SM}}}
\times[(a_1+1)^2\sigma_{\text{SM}}^H-\sigma_{\text {SM}}^H+a^2_2\sigma^H_{CP-{\text{even}}}
+a^2_3\sigma^H_{CP-{\text{odd}}}+(a_1+1)a_2\sigma_{CP-{\text{even}},{\text {SM}}}^{\rm int}]\\
&&+
\frac{N_{gg~~{\rm interference}}}{\sigma^{\rm int}_{\text{SM}}}
\times[a_1\sigma_{{\rm SM},{\rm box}}^{\rm int}+a_2\sigma^{\rm int}_{CP-{\text{even},{\rm box}}}],
\label{Nexp}
\eea
where $N^{\rm exp}(a_1,a_2,a_3)$ represents the expected number of events
from anomalous $CP$-even and $CP$-odd processes,
$\sigma_k^H$ is the self-conjugate Higgs-mediated cross section, and $\sigma_{k,l}^{\rm int}$ is the interference cross section with $k,l=$~\{box, SM, $CP$-even, $CP$-odd\}.
The first term on the right-hand side of the equation is the contribution from the s-channel processes, and the second part is the contribution from the interference between the s-channel processes and the box diagram. For each category with the same topological Feynman diagrams, it is assumed to have the same  $k$-factor and total efficiency $\epsilon$, which are equal to the corresponding values for the SM process. These coefficients are extracted from experimental measurements, which are similar as the treatment in the experiments~\cite{Sirunyan:2019twz,Ellis:2014yca}.

The cross section of $4\ell$ final states is the sum of
the cross sections of $2e2\mu$, $4e$ and $4\mu$ final states.
$N^{\rm exp}(a_1,a_2,a_3)$ can be obtained by combining the corresponding cross sections from both Table.~\ref{table:13tevsigma}
and Table.~\ref{table:13tevsigma_4l}.

The experimental observed number $N^{\rm obs}(a_1,a_2,a_3)$ that corresponds to $N^{\rm exp}(a_1,a_2,a_3)$ is defined as
$N^{\rm obs}(a_1,a_2,a_3)=N_{\rm total~~observed }-N^{\rm SM}_{\rm total~~expected }=38.7$ in the CMS experiment\cite{Sirunyan:2019twz}. Its fluctuation is estimated as the $\delta_{\rm off-shell}=\sqrt{N_{\text{total observed}}}=\sqrt{1325}$~(including both signal and background).

Second, the observed signal strength of the $gg\to H\to 4l$ process measured by CMS~\cite{CMS-PAS-HIG-19-001} is $\mu_{ggH}^{\rm obs}=0.97^{+0.09}_{-0.09}\mbox{(stat.)}^{+0.09}_{-0.07}\mbox{(syst.)}$. Its fluctuation is $\delta_{\rm on-shell}=0.127   $ after a combination of both statistical and systematic errors.  Theoretically, the signal strength with anomalous couplings can be estimated as
\beq
\mu_{ggH}^{\rm{exp}}(a_1,a_2,a_3)=\frac{1}{\sigma_{\rm SM}^H}[(a_1+1)^2\sigma_{\rm SM}^H+a_2^2\sigma_{CP-{\rm even}}^H+a_3^2\sigma_{CP-{\rm odd}}^H+(a_1+1)a_2\sigma_{CP-{\rm even},{\rm SM}}^{\rm int}],
\label{mu}
\eeq
where $\sigma_k^H$ and $\sigma_{k,l}^{\rm int}$ are same as in Eq.(\ref{Nexp}) except in the on-shell region.
Equation~(\ref{mu}) is shorter than Eq.(\ref{Nexp}) because in the on-shell Higgs region
the interference term with box diagram $\sigma_{\rm SM, box}$ and $\sigma_{\rm CP-even, box}$ are zero.

The survival parameter regions of $a_1,a_2$ and $a_3$ can be obtained by a global $\chi^2$ fit, which can be constructed as
\beq
\chi^2=\left(\frac{N^{\rm exp}-N^{\rm obs}}{\delta_{\rm off-shell}}\right)^2+\left(\frac{\mu_{ggH}^{\rm exp}-\mu_{ggH}^{\rm{obs}}}{\delta_{\rm on-shell}}\right)^2.
\label{chi2}
\eeq

The adoption of the $\chi^2$ fit here can be controversial, as we only have two input data points (on-shell and off-shell) and have to find parameter regions for three variables ($a_1,a_2$ and $a_3$). We claim that the result here is just for a complete analysis including both theoretical calculation and experimental constraints and it is very preliminary. The situation can be improved if experimental collaborations can collect sufficient statistics in the future. Nevertheless, the $\chi^2$ fit can also provide some interesting results.

\begin{figure}[htbp]
\centering
\centerline{\includegraphics[height=5cm,width=5cm]{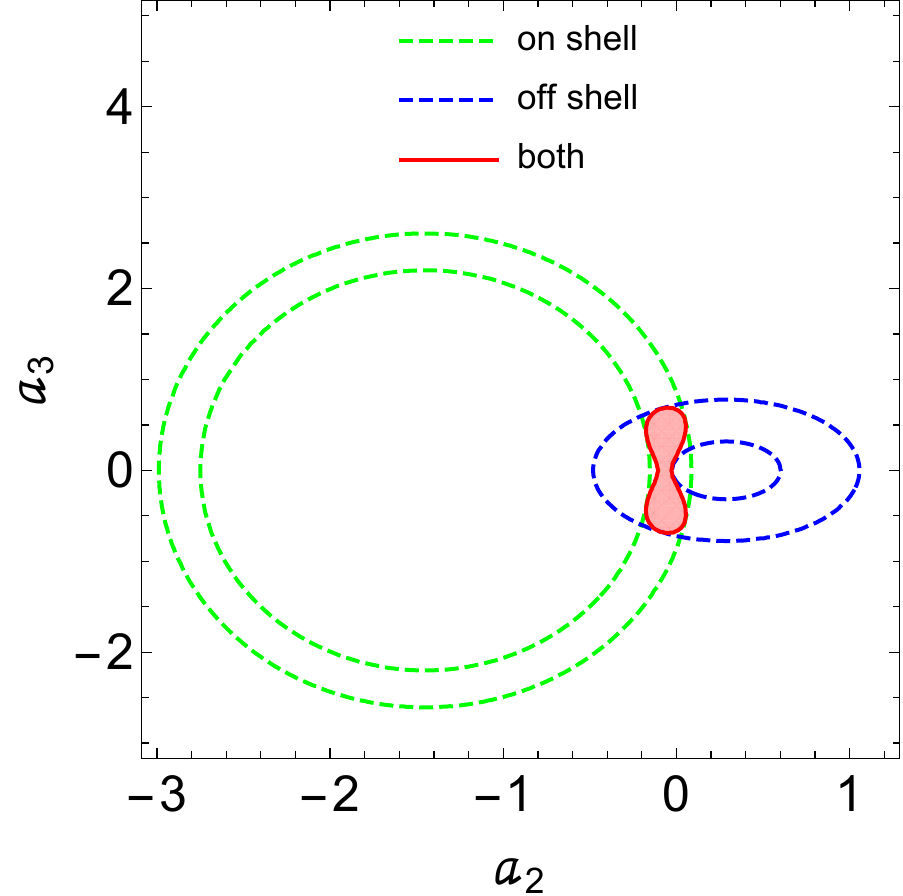}
\includegraphics[height=5cm,width=5cm]{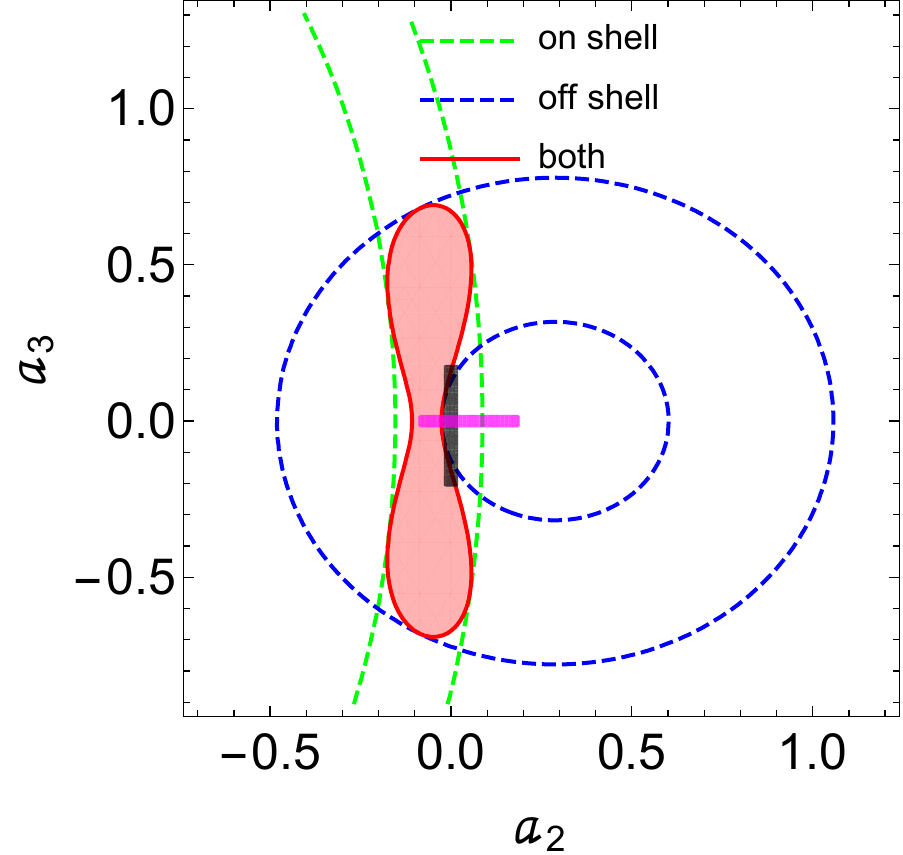}
}
\centerline{\includegraphics[height=5cm,width=5cm]{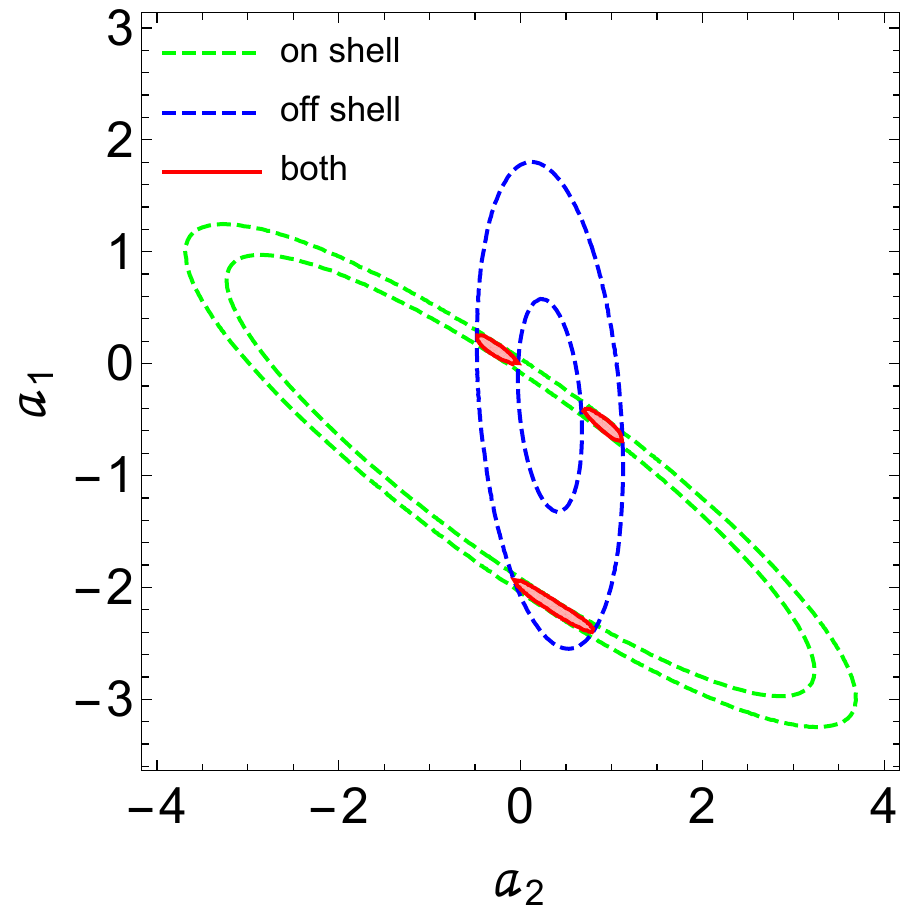}
\includegraphics[height=5cm,width=5cm]{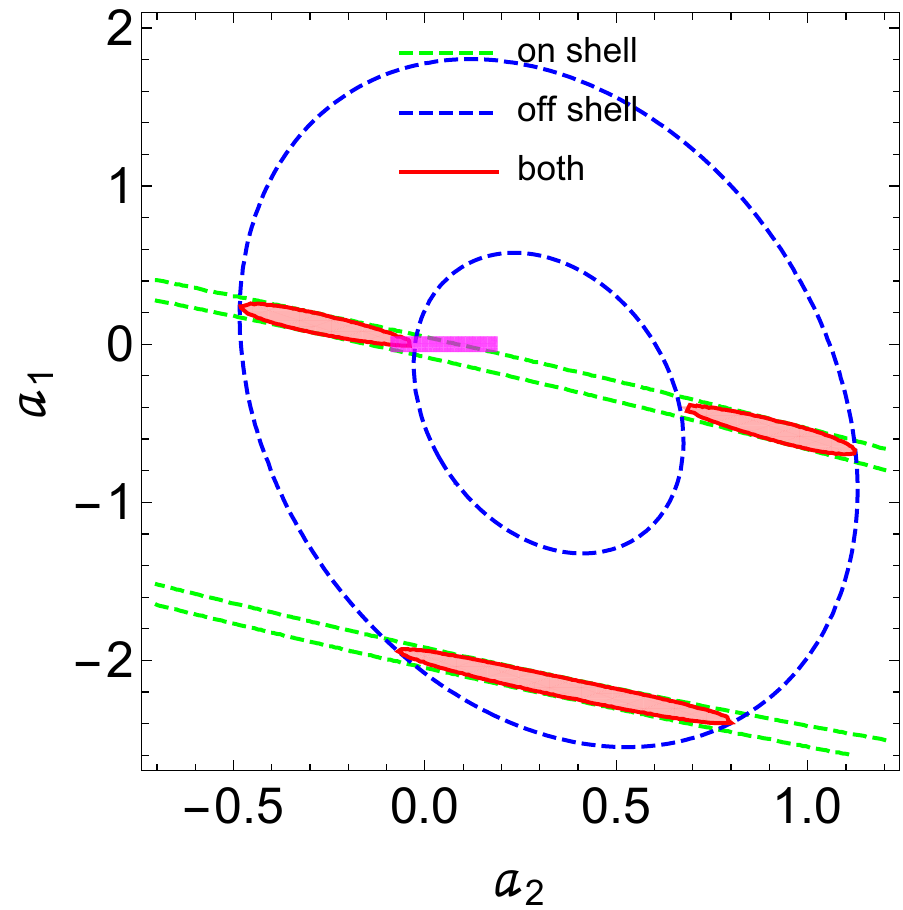}}
\centerline{\includegraphics[height=5cm,width=5cm]{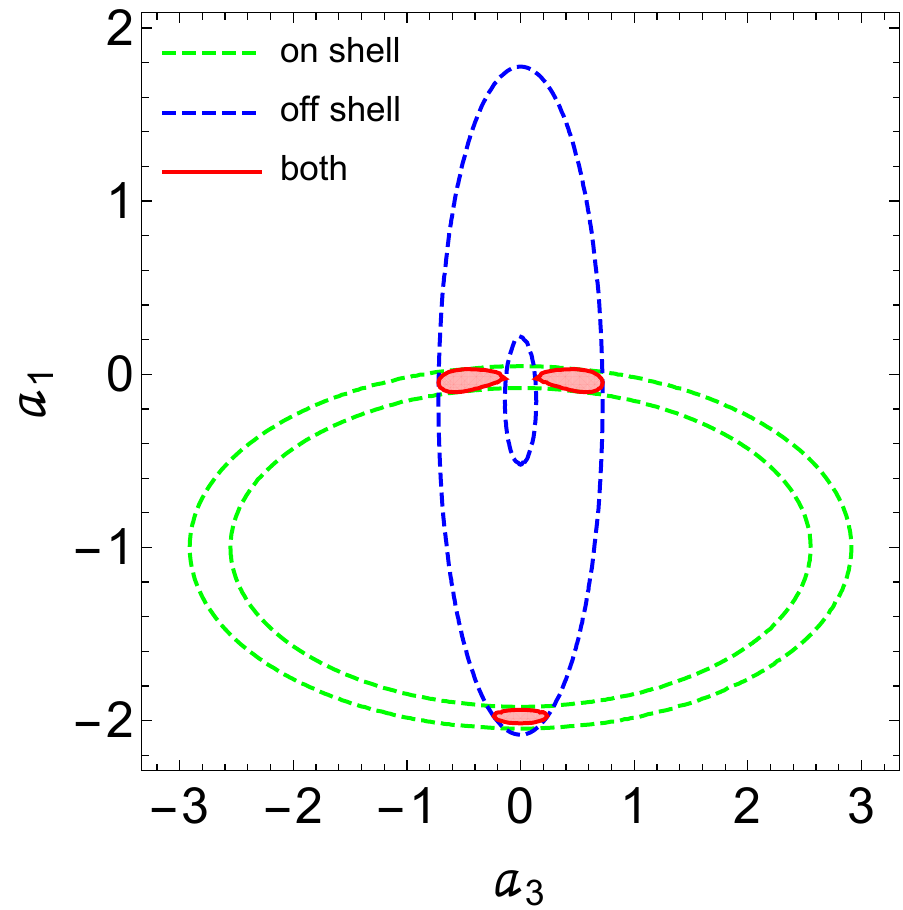}
\includegraphics[height=5cm,width=5cm]{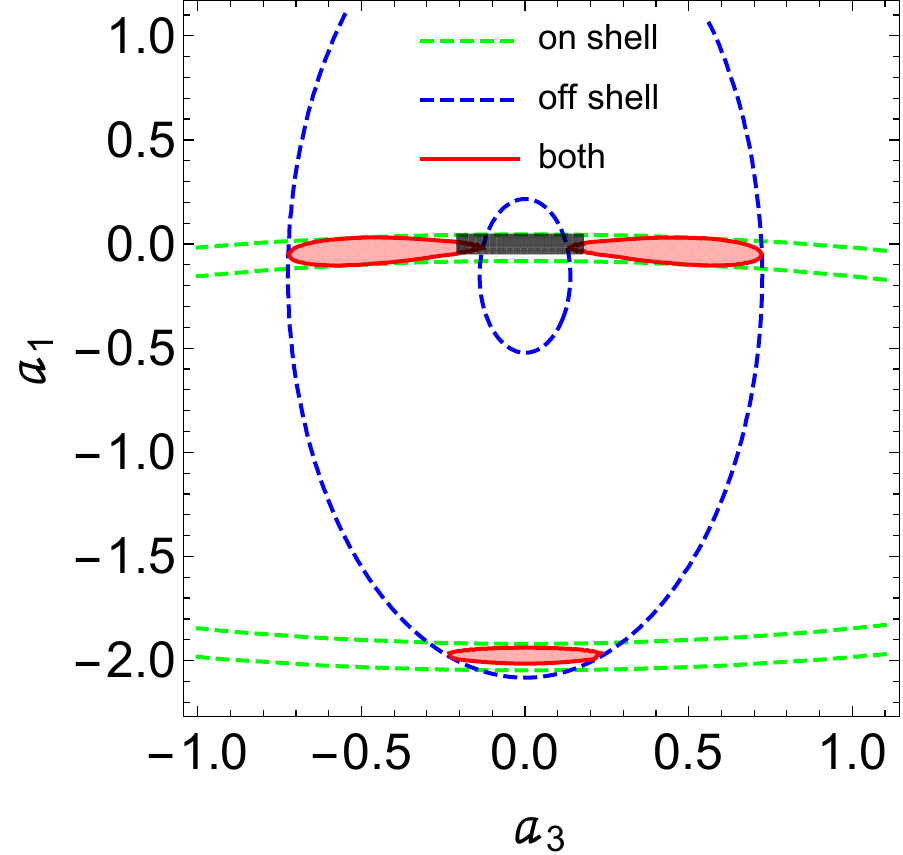}}
\caption{Two-dimensional constraints on the new physics coefficients $a_1$,$a_2$ and $a_3$ from $\chi^2$ fits. To illustrate the constraints from different energy regions, three $1\sigma$ regions (green concentric circles, blue concentric circles, and red region) from three individual $\chi^2$ fits (on-shell, off-shell, and both) are drawn here. CMS $2\sigma$ constraints~($95\%$ confidence level)~\cite{Sirunyan:2019twz} are drawn as the lines~(magenta for $a_2$ when $a_1=a_3=0$ and grey for $a_3$ when $a_1=a_2=0$) in the right zoom in plots.}
\label{contour}
\end{figure}

Fig.\ref{contour} shows the two dimensional contour diagram of the anomalous couplings.
There are three colored regions~(green, blue, and red) in each small plot and the red areas are the final $1\sigma$ survival parameter regions from the global $\chi^2$ fit. In the actual two dimensional fitting procedure, we take two anomalous couplings to be free and fix the third one to be zero. Three individual $\chi^2$ fits are operated, constraint only from the off-region (first part in Eq.~\eqref{chi2}), constraint from the on-shell region~(second part in Eq.~\eqref{chi2}) and both of the two. The purpose is to show how the irregular overlap red regions come from. As discussed in above sections, we have equal number of experimental data points and free parameters here and the $\chi^2$ fit degenerates to an equation solving problem. Survival parameter regions from either on-shell or off-shell constraint come to be concentric circles and the global fitting results are almost the overlap region between them.

In the recently updated CMS experiment~\cite{Sirunyan:2019twz},
it use both on-shell and off-shell data, construct
kinematic discriminants, and get the limit (at 95\% confidence level) of the parameters $a_2\subset [-0.09,0.19]$,~$a_3\subset [-0.21,0.18]$~(there is no corresponding constraint on $a_1$). This experimental analysis is based on one free parameter-fitting schedule so we draw them as the line segments in the right plots of Fig.~\ref{contour}~(magenta for $a_2$ and grey for $a_3$). Our global fit results is roughly consistent with the CMS's, although within a first glance the two seems have some tension(Pay attention that we draw $1\sigma$ contour while CMS's results are the limit at 95\% confidence level which corresponds to $2\sigma$ intervals in the hypothesis of Gaussian distribution). The CMS's results seems to be more stringent than ours. This maybe caused by more kinematic information in detail they used in their analysis. Besides, we have some parameter regions with $a_1\sim -2$ or $a_2$ approaching 1. These regions show the correlations of each pairs of parameters. There is cancellation on the cross sections when the parameters coexist. In principle, the anomalous couplings should be much smaller than 1 to validate the operator expansion. Therefore, these parameter regions should be ruled out. Nevertheless, our global fit provides a complementary perspective of how the final anomalous coupling parameters contour regions are obtained from the individual on-shell/off-shell energy region constraints. These preliminary fitting results can be optimized in the case of more statistics in the future.

\section{conclusion and discussion\label{section:conclusion}}

When considering the anomalous $HZZ$ couplings, we calculate
the cross sections induced by these new couplings,
and special attention is focused on the interference effects.
In principle, there are three kinds of interference:
1. the interference between anomalous $CP$-even Higgs-mediated process and the continuum background box process $\sigma_{CP\text{-even},\text{box}}$;
2. the interference between anomalous $CP$-even Higgs-mediated process and SM Higgs-mediated process $\sigma_{CP\text{-even},\text{SM}}$;
and 3. the interference between the anomalous $CP$-odd Higgs-mediated process
and all other processes $\sigma_{CP\text{-odd},k}$ with $k = {\text{box}, \text{SM},
CP\text{-even}}$.
The numerical results of the integrated cross sections show that
the first kind of interference can be neglected in the on-shell Higgs region but is nonnegligible in the off-shell Higgs region,
the second kind of interference
is important in both the on-shell and off-shell Higgs regions,
 and the third kind of interference has zero contribution for the total cross section in both regions.

By using the theoretical calculation together with both on-shell and off-shell Higgs experimental data, 
we estimate the constraints on the anomalous $HZZ$ couplings. 
The correlations of the different kinds of anomalous couplings
are shown in contour plots, which illustrate how the
anomalous contributions cancel each other out and the extra parameter 
regions survive when they coexist.   


In this research we only use the numerical results of integrated
cross sections, whereas in fact more information can be fetched from the
differential cross sections (kinematic distributions).
Furthermore, the $k$-factors and total efficiencies
should also be estimated separately according to different sources.
We leave them for our future work.

\begin{acknowledgements}
We thank John M. Campbell for his helpful explanation of the code in the \texttt{MCFM} package. The work is supported by the National Natural Science Foundation of China under Grant No.11847168, the Fundamental Research Funds for the Central Universities of China under Grant No. GK201803019, GK202003018, 1301031995, and the Natural Science Foundation of Shannxi Province, China (2019JM-431, 2019JQ-739).

\end{acknowledgements}

\begin{appendix}

\section{Helicity amplitudes for the process $H\rightarrow ZZ \to e^-e^+\mu^-\mu^+$\label{app:amp}}

The helicity amplitudes $\mathcal{A}_1$, $\mathcal{A}_2$ and $\mathcal{A}_3$ are shown separately.
The common factor $f$ is defined as
\beq
f=-2ie^3\frac{1}{M_W \sin\theta_W}\frac{P_Z(s_{34})}{s_{34}}\frac{P_Z(s_{56})}{s_{56}}~. \nonumber
\eeq

\beq
\begin{aligned}
& \mathcal{A}_1^{H\rightarrow ZZ\to2e2\mu}(3^-_{e^-},4^+_{e^+},5^-_{\mu^-},6^+_{\mu^+})=
f \times  l^2_e\frac{M_W^2}{\cos^2\theta_W}\langle35\rangle[46] , \\
& \mathcal{A}_1^{H\rightarrow ZZ\to2e2\mu}(3^-_{e^-},4^+_{e^+},5^+_{\mu^-},6^-_{\mu^+})=
f \times  l_er_e\frac{M_W^2}{\cos^2\theta_W}\langle36\rangle[45] , \\
& \mathcal{A}_1^{H\rightarrow ZZ\to2e2\mu}(3^+_{e^-},4^-_{e^+},5^-_{\mu^-},6^+_{\mu^+})=
f \times  l_er_e\frac{M_W^2}{\cos^2\theta_W}\langle45\rangle[36] , \\
& \mathcal{A}_1^{H\rightarrow ZZ\to2e2\mu}(3^+_{e^-},4^-_{e^+},5^+_{\mu^-},6^-_{\mu^+})=
f \times  r^2_e\frac{M_W^2}{\cos^2\theta_W}\langle46\rangle[35] ~.
\end{aligned}
\eeq

\beq
\begin{aligned}
&
\mathcal{A}_2^{H\rightarrow ZZ\to2e2\mu}(3^-_{e^-},4^+_{e^+},5^-_{\mu^-},6^+_{\mu^+})=
f\times l^2_e\times \\
&
\Big[2k\cdot k^{\prime}\langle35\rangle[46]
 +\big(\langle35\rangle[45]+
\langle36\rangle[46]\big)\big(\langle35\rangle[36]+\langle45\rangle[46]\big) \Big],  \\
&
\mathcal{A}_2^{H\rightarrow ZZ\to2e2\mu}(3^-_{e^-},4^+_{e^+},5^+_{\mu^-},6^-_{\mu^+})=
f\times l_er_e\times \\
&
\Big[2k\cdot k^{\prime}\langle36\rangle[45]
 +\big(\langle35\rangle[45]+
\langle36\rangle[46]\big)\big(\langle36\rangle[35]+\langle46\rangle[45]\big) \Big],  \\
&
\mathcal{A}_2^{H\rightarrow ZZ\to2e2\mu}(3^+_{e^-},4^-_{e^+},5^-_{\mu^-},6^+_{\mu^+})=
f\times r_el_e\times \\
&
\Big[2k\cdot k^{\prime}\langle45\rangle[36]
 +\big(\langle45\rangle[35]+
\langle46\rangle[36]\big)\big(\langle35\rangle[36]+\langle45\rangle[46]\big) \Big],  \\
&
\mathcal{A}_2^{H\rightarrow ZZ\to2e2\mu}(3^+_{e^-},4^-_{e^+},5^+_{\mu^-},6^-_{\mu^+})=
f\times r^2_e\times \\
&
\Big[2k\cdot k^{\prime}\langle46\rangle[35]
 +\big(\langle45\rangle[35]+
\langle46\rangle[36]\big)\big(\langle36\rangle[35]+\langle46\rangle[45]\big) \Big].
\end{aligned}
\eeq

\beq
\begin{aligned}
&
\mathcal{A}_3^{H\rightarrow ZZ\to2e2\mu}(3^-_{e^-},4^+_{e^+},5^-_{\mu^-},6^+_{\mu^+})=
f\times l^2_e\times (-i) \times \\
& \Big[2\big(k\cdot k^{\prime}+\langle46\rangle[46]\big)\langle35\rangle[46]
+\langle35\rangle[45]\big(\langle35\rangle[36]+\langle45\rangle[46]\big)      \\
&+\langle36\rangle[46]\big(\langle35\rangle[36]-\langle45\rangle[46]\big)\Big]~,  \\
&
 \mathcal{A}_3^{H\rightarrow ZZ\to2e2\mu}(3^-_{e^-},4^+_{e^+},5^+_{\mu^-},6^-_{\mu^+})=
f\times l_er_e\times (-i) \times \\
& \Big[2\big(k\cdot k^{\prime}+\langle45\rangle[45]\big)\langle36\rangle[45]
+\langle36\rangle[46]\big(\langle36\rangle[35]+\langle46\rangle[45]\big)      \\
&+\langle35\rangle[45]\big(\langle36\rangle[35]-\langle46\rangle[45]\big)\Big],  \\
&
 \mathcal{A}_3^{H\rightarrow ZZ\to2e2\mu}(3^+_{e^-},4^-_{e^+},5^-_{\mu^-},6^+_{\mu^+})=
f\times r_el_e\times (-i) \times  \\
& \Big[2\big(k\cdot k^{\prime}+\langle36\rangle[36]\big)\langle45\rangle[36]
+\langle45\rangle[35]\big(\langle45\rangle[46]+\langle35\rangle[36]\big)      \\
&+\langle46\rangle[36]\big(\langle45\rangle[46]-\langle35\rangle[36]\big)\Big], \\
&
 \mathcal{A}_3^{H\rightarrow ZZ\to2e2\mu}(3^+_{e^-},4^-_{e^+},5^+_{\mu^-},6^-_{\mu^+})=
f\times r^2_e\times (-i) \times \\
& \Big[2\big(k\cdot k^{\prime}+\langle35\rangle[35]\big)\langle46\rangle[35]
+\langle46\rangle[36]\big(\langle46\rangle[45]+\langle36\rangle[35]\big)      \\
&+\langle45\rangle[35]\big(\langle46\rangle[45]-\langle36\rangle[35]\big)\Big].
\end{aligned}
\eeq

\section{The cross sections at $\sqrt{s}=8$~TeV
\label{table8tev}}

\begin{table*}[!htp]
\begin{floatrow}

\begin{minipage}{0.5\linewidth}
\centering

\begin{tabular}{|c|c|c|c|c|c|}
 \hline
 \multicolumn{6}{|c|}{8~\text{TeV},~$m_{2e2\mu}<130\rm~GeV$, on-shell} \\
 \hline
 \multicolumn{2}{|c|}{\multirow{2}{*}{$\sigma_{k,l}$(fb)}} & \multirow{2}{*}{box} &
\multicolumn{3}{c|}{\footnotesize Higgs-med.}\\
 \cline{4-6}
 \multicolumn{2}{|c|}{} & & {\footnotesize SM} & {\footnotesize $CP$-even} & {\footnotesize $CP$-odd}  \\
\hline
\multicolumn{2}{|c|}{box} & 0.011 & 0 & 0 & 0 \\
\hline
\multirow{3}{*}{ \begin{tabular}{c} \rotatebox{90}{\footnotesize Higgs-med.} \end{tabular}} &
 {\footnotesize SM} & 0 & 0.232 & 0.257 & 0 \\
\cline{2-6}
 & \footnotesize $CP$-even & 0 & 0.257 & 0.093 & 0 \\
\cline{2-6}
 & \footnotesize $CP$-odd & 0 & 0 & 0 & 0.035 \\
\hline
 \end{tabular}

\end{minipage}

\hfill

\begin{minipage}{0.5\linewidth}

\begin{tabular}{|c|c|c|c|c|c|}
 \hline
 \multicolumn{6}{|c|}{8~\text{TeV}~,~$m_{2e2\mu}>220\rm~GeV$, off-shell} \\
 \hline
 \multicolumn{2}{|c|}{\multirow{2}{*}{$\sigma_{k,l}$(fb)}} & \multirow{2}{*}{box} &
\multicolumn{3}{c|}{\footnotesize Higgs-med.}\\
 \cline{4-6}
 \multicolumn{2}{|c|}{} & & {\footnotesize SM} & {\footnotesize $CP$-even} & {\footnotesize $CP$-odd}  \\
\hline
\multicolumn{2}{|c|}{box} & 0.479 & -0.056 & -0.198 & 0 \\
\hline
\multirow{3}{*}{ \begin{tabular}{c} \rotatebox{90}{\footnotesize Higgs-med.} \end{tabular}} &
 {\footnotesize SM} & -0.056 & 0.031 & 0.047 & 0 \\
\cline{2-6}
 & \footnotesize $CP$-even & -0.198 & 0.047 &  0.228 & 0 \\
\cline{2-6}
 & \footnotesize $CP$-odd & 0 & 0 & 0 & 0.219 \\
\hline
 \end{tabular}

\end{minipage}

\caption{ Cross sections of $gg\to2e2\mu$ process in proton-proton collision at $\sqrt{s}=8$~TeV with $a_1=0,a_2=a_3=1$ in Eq.~\eqref{eqn:gamma}.}
\label{table:8tevsigma}
\end{floatrow}
\end{table*}

\begin{table*}[!htp]
\begin{floatrow}

\begin{minipage}{0.5\linewidth}
\centering

\begin{tabular}{|c|c|c|c|c|c|}
 \hline
 \multicolumn{6}{|c|}{8~\text{TeV}~,~$m_{4e/4\mu}<130\rm~GeV$, on-shell} \\
 \hline
 \multicolumn{2}{|c|}{\multirow{2}{*}{$\sigma_{k,l}$(fb)}} & \multirow{2}{*}{box} &
\multicolumn{3}{c|}{\footnotesize Higgs-med.}\\
 \cline{4-6}
 \multicolumn{2}{|c|}{} & & {\footnotesize SM} & {\footnotesize $CP$-even} & {\footnotesize $CP$-odd}  \\
\hline
\multicolumn{2}{|c|}{box} & 0.021 & 0 & 0 & 0 \\
\hline
\multirow{3}{*}{ \begin{tabular}{c} \rotatebox{90}{\footnotesize Higgs-med.} \end{tabular}} &
 {\footnotesize SM} & 0 & 0.248 & 0.261 & 0 \\
\cline{2-6}
 & \footnotesize $CP$-even & 0 & 0.261 & 0.086 & 0 \\
\cline{2-6}
 & \footnotesize $CP$-odd & 0 & 0  & 0 & 0.028  \\
\hline
 \end{tabular}

\end{minipage}

\hfill

\begin{minipage}{0.5\linewidth}

\begin{tabular}{|c|c|c|c|c|c|}
 \hline
 \multicolumn{6}{|c|}{8~\text{TeV}~,~$m_{4e/4\mu}>220\rm~GeV$, off-shell} \\
 \hline
 \multicolumn{2}{|c|}{\multirow{2}{*}{$\sigma_{k,l}$(fb)}} & \multirow{2}{*}{box} &
\multicolumn{3}{c|}{\footnotesize Higgs-med.}\\
 \cline{4-6}
 \multicolumn{2}{|c|}{} & & {\footnotesize SM} & {\footnotesize $CP$-even} & {\footnotesize $CP$-odd}  \\
\hline
\multicolumn{2}{|c|}{box} & 0.485 & -0.056 & -0.199 & 0 \\
\hline
\multirow{3}{*}{ \begin{tabular}{c} \rotatebox{90}{\footnotesize Higgs-med.} \end{tabular}} &
 {\footnotesize SM} & -0.056 & 0.031 & 0.047 & 0 \\
\cline{2-6}
 & \footnotesize $CP$-even & -0.199 & 0.047 &  0.229 & 0  \\
\cline{2-6}
 & \footnotesize $CP$-odd & 0 & 0 & 0 & 0.215 \\
\hline
 \end{tabular}

\end{minipage}

\caption{ The cross sections of $gg\to4e/4\mu$ processes in proton-proton collision at
center-of-mass energy $\sqrt{s}=8$~TeV with $a_1=0, a_2=a_3=1$ in Eq.~\eqref{eqn:gamma}.
}
\label{table:8tevsigma_4l}
\end{floatrow}
\end{table*}

\end{appendix}

\bibliographystyle{utphys} 
\bibliography{reference}

\end{document}